\documentclass[twocolumn]{aastex6}
\usepackage{times}
\usepackage{amsmath}
\usepackage{textcomp}
\usepackage{amssymb}
\usepackage[flushleft]{threeparttable}
\usepackage{boldline}
\usepackage{tabularx}
\usepackage{lineno}

\newcommand{\lum}{erg\,s$^{-1} $}
\newcommand{\fermi}{{\it Fermi}}
\newcommand{\nustar}{{\it NuSTAR}}

\newcommand{\swift}{{\it Swift}}

\newcommand{\ergflux}{\mbox{${\rm \, erg \,\, cm^{-2} \, s^{-1}}$}}
\newcommand{\gm}{$\gamma$}

\slugcomment{ApJ submitted}

\shorttitle{MeV \nustar~Blazars}
\shortauthors{Marcotulli et al.}

\begin{document}

\title{High-redshift blazars through \nustar~eyes}

\author{L. Marcotulli$^{1}$,  V. S. Paliya$^{1}$, M. Ajello$^{1}$, A. Kaur$^{1}$, D. H. Hartmann$^{1}$, D. Gasparrini$^{2,3}$, J.Greiner$^{4}$, A. Rau$^{4}$, P. Schady$^{4}$, M. Balokovi\'c$^{5}$, D. Stern$^{6}$, G. Madejski$^{7}$} 
\affil{$^1$Department of Physics and Astronomy, Clemson University, Kinard Lab of Physics, Clemson, SC 29634-0978, USA}
\affil{$^2$Agenzia Spaziale Italiana (ASI) Science Data Center, I-00133 Roma, Italy}
\affil{$^3$Istituto Nazionale di Fisica Nucleare, Sezione di Perugia, I-06123 Perugia, Italy}
\affil{$^{4}$Max Planck Institute f\"ur extraterrestrische Physik, Giessenbachstrasse 1, 85748, Garching, Germany}
\affil{$^{5}$Cahill Center for Astronomy and Astrophysics, California Institute of Technology, Pasadena, CA 91125,USA}
\affil{$^{6}$Jet Propulsion Laboratory, California Institute of Technology, Pasadena, CA 91109, USA}
\affil{$^{7}$Kavli Institute for Particle Astrophysics and Cosmology, SLAC National Accelerator Laboratory, Menlo Park, CA 94025, USA}
\email{lmarcot@g.clemson.edu, vpaliya@g.clemson.edu}

\begin{abstract}
The most powerful sources among the blazar family are MeV blazars. Often detected at $z>2$, they usually display high X- and \gm-ray luminosities, larger-than-average jet powers and black hole masses $\gtrsim 10^9 M_{\sun}$. 
In the present work we perform a multiwavelength study of three high redshift blazars: 3FGL J0325.5+2223 ($z=2.06$),  3FGL J0449.0+1121 ($z= 2.15$), and 3FGL J0453.2$-$2808 ($z=2.56$), analysing quasi simultaneous data from
GROND, \swift-UVOT and XRT, \nustar, and \fermi-LAT. Our main focus is on the hard X-ray  band recently unveiled by \nustar~(3$-$79 keV) where these objects show a hard 
spectrum which enables us to constrain the inverse Compton peak and the jet power. We found that all three targets resemble the most powerful blazars, with the synchrotron peak located in the sub-millimeter range and the inverse Compton peak in the MeV range, and therefore 
belong to the MeV blazar class.
Using a simple one zone leptonic emission model to reproduce the spectral energy distributions, we conclude that a simple combination of synchrotron and accretion disk emission reproduces
the infrared-optical spectra while the X-ray to \gm-ray part is well reproduced by the inverse Compton scattering of low energy photons supplied
by the broad line region. The black hole masses for each of the three sources are calculated to be $\gtrsim 4 \times 10^{8} M_{\sun}$. The three studied sources have jet power at the level of, or beyond, the accretion luminosity.

\end{abstract}


\section{Introduction}\label{sec:intro}

Blazars are a subset of active galactic nuclei (AGN) whose relativistic jets are pointed towards
the observer ($\theta_V < 1/\Gamma$, $\theta_V$ being the viewing angle and $\Gamma$ the bulk Lorentz factor).
From their optical properties, they have been sub-classified as flat spectrum radio quasars (FSRQs), if their spectrum
shows broad emission lines (equivalent width, EW$>5$\AA), and BL Lacertae (BL Lac) objects which show weak (EW$<5$\AA) or 
absent emission lines 
\citep{1991ApJ...374..431S,1991ApJS...76..813S}. Blazars are known to radiate over the entire electromagnetic spectrum, 
from the low-energy radio band to very high-energy \gm-rays; this radiation is 
primarily due to non-thermal emission processes and is believed to be the manifestation of a powerful relativistic 
jet \citep[e.g.,][]{1978PhyS...17..265B,1995PASP..107..803U,2016CRPhy..17..594D}.
The spectral energy distribution (SED) of a blazar displays two characteristic broad humps, one peaking between
infrared (IR) and X-ray frequencies and the other
between the X-ray and \gm-ray energy bands. The low-energy hump is attributed to the synchrotron process of relativistic 
electrons present in the jet,
while the high-energy peak is associated with the inverse Compton (IC) scattering of low-energy photons
by relativistic electrons. The low-energy photons can be either
synchrotron photons (synchrotron self Compton or SSC; \citealp{1981ApJ...243..700K}) or photons originating
externally to the jet (external Compton or EC; \citealp{1987ApJ...322..650B}).

Two features are noteworthy: due to the peculiar orientation of the relativistic jet, the enhancement ascribed to relativistic beaming allows us
to detect blazars at high redshifts. Moreover, according to the so-called `blazar sequence' \citep{1998MNRAS.299..433F,1998MNRAS.301..451G},
the efficiency of electron cooling due to IC increases with increasing source luminosity. These arguments imply that
the more distant and luminous the object is, the greater the shift of the SED humps towards lower frequencies. 
The synchrotron peak of the most powerful blazars is located in the sub-millimeter (mm) range,
while the IC peak falls in the MeV band.
The position of the latter classifies such objects as `MeV blazars' \citep[e.g.,][]{1995A&A...293L...1B}. 
The characteristics of their spectra have allowed us to detect blazars up to $z\geq5$ \citep[e.g.,][]{2004ApJ...610L...9R,2013ApJ...777..147S}.

The shift of the IC peak makes MeV blazars bright at hard X-rays 
($>10$ keV) \citep[see, e.g.,][]{2013MmSAI..84..719G}.
With the advent of the first focusing hard X-ray telescope in orbit, the {\it Nuclear Spectroscopic Telescope Array} (\nustar; \citealp{2013ApJ...770..103H}), 
it is now possible to explore the hard X-ray (3$-$79 keV) energy band with unprecedented detail. 
\nustar~ has already been shown to be a powerful instrument for the study of the most 
luminous and distant blazars, opening a window to understand the early X-ray universe \citep[e.g.,][]{2013ApJ...777..147S}. 
Thanks to its sensitivity, it has allowed us to investigate some  peculiar X-ray  features of high-redshift blazars,
like variability \citep[e.g.][]{2016MNRAS.462.1542S} or flattening in the spectrum \citep[e.g.,][]{2016ApJ...825...74P}.
Also, by combining \nustar~observations with \fermi-Large Area Telescope \citep[\fermi-LAT;][]{2009ApJ...697.1071A} data, we can more reliably measure the location of the IC peak in the SED. Since the bolometric luminosity of blazars is dominated  
by high-energy emission \citep[e.g.,][]{2014Natur.515..376G}, an accurate measurement of the IC component in the SED provides important constraints on the power of the jet
and on the relativistic particle population.
Moreover, the shift in the synchrotron peak reveals the underlying optical-UV thermal emission from the accretion disk \citep[e.g.,][]{2010MNRAS.405..387G}. Modeling the disk emission  
with a standard optically thick, spatially thin geometry \citep[][]{1973A&A....24..337S}, one can estimate the central black hole mass and the accretion disk luminosity. As a result, 
MeV blazars are ideal objects to study the accretion-jet connection.
Furthermore, these sources generally host extremely massive black holes at their centers  \citep[$M_{\rm BH}\gtrsim 10^9 M_{\sun}$, e.g.,][]{2010MNRAS.405..387G,2016ApJ...826...76A,2016ApJ...825...74P}. 
This has important implications as the detection of a blazar implies the existence of $2\Gamma^2$ such sources (where $\Gamma\sim$10-15, e.g., \citealp{1997ApJ...484..108S}) 
with misaligned jets at the same redshift, hosting similarly massive black holes. Therefore a detailed study of MeV blazars, hosting extremely massive black holes, places useful 
constraints on the high end of the black-hole mass function, which is essential for a full theoretical understanding of the growth and evolution of black holes over cosmic time (e.g., \citealp{2016PASA...33....7J}).
This can be accomplished by adopting a multi-wavelength approach and utilizing data from a variety of
instruments. \\

Here we present a broadband study of three high-redshift blazars: 3FGL J0325.5$+$2223 ($z=2.06$),  3FGL J0449.0$+$1121 ($z= 2.15$), and 3FGL J0453.2$-$2808 ($z=2.56$).
These are among the most powerful known sources of this class: found at $z>2$ with a \gm-ray luminosity L$_{\gamma}>10^{46}$ \lum \citep[][]{2015ApJ...810...14A}, 
they are soft \gm-ray emitters
but display a hard X-ray continuum (2-10 keV) \citep[e.g.,][]{2009ApJ...699..603A,2011MNRAS.411..901G}, which classifies them as MeV blazars.
They are among the few of their class that are detected in X-rays and by the \fermi-LAT as well. 

In fact, they have been selected as they are the only three sources, among the ten most luminous LAT FSRQs \citep[][]{2015ApJ...810...14A},
which do not have hard X-ray coverage. 
Therefore we report their first E$>$10 keV detection obtained from \nustar~observations\footnote{These three sources were observed by
\nustar~ as part of our cycle program (proposal number 1285; obs IDs: 60101078002, 60101079002, 60101080002).}. As such it becomes possible to
accurately determine the location of the high-energy peak in their SEDs.

Our primary motivation is to understand their physical properties by means of a multi-frequency data analysis and theoretical SED modeling, with a major focus on the hard X-ray observations. 
All three sources have been simultaneously observed by \nustar, \swift~X-Ray Telescope \citep[XRT;][]{2005SSRv..120..165B} 
and \swift-UltraViolet and Optical Telescope \citep[UVOT;][]{2005SSRv..120...95R}. Thus, the X-ray energy band was fully covered 
from 0.3 up to 79 keV. To cover the infrared (IR) to ultra-violet (UV) part of the SEDs, we integrated the observations from 
\swift-UVOT with the ones from the Gamma-Ray Burst Optical/Near-Infrared Detector \citep[GROND;][]{2008PASP..120..405G}; for two of the sources these were
carried out within one week of \nustar~ observations, while for 3FGL J0453.2$-$2808 they were taken within six months, due to technical issues. 
We also analyze the recently released Pass 8 data from \fermi-LAT, which provides better sensitivity at lower energies \citep[][]{2013arXiv1303.3514A} compared to previously released datasets.
Throughout, we use cosmological 
parameters $H_0=71$~km~s$^{-1}$~Mpc$^{-1}$, $\Omega_m = 0.27$, and $\Omega_\Lambda = 0.73$ \citep{2009ApJS..180..330K}.

\section{Observations}

\subsection{\fermi}
The LAT Pass 8 data used in this work covers the period of \nustar~observations. Since all three objects are faint in \gm-rays, we chose a 
large time bin (MJD 57082$-$57448) to generate a meaningful SED. Moreover, there is no significant \gm-ray variability detected from these 
sources\footnote{We searched for significant \gm-ray flux variations using the tool `\fermi~ All-sky Variability Analysis' \citep[FAVA;][]{2013ApJ...771...57A},
but found none, at least during the period covered in this work.}, and therefore, the selected period is a reasonable choice.

We followed the standard data reduction procedure as given in the online documentation\footnote{http://fermi.gsfc.nasa.gov/ssc/data/analysis/documentation/} with a
few modifications. In the energy range 0.06$-$300 GeV, we only selected SOURCE class events ({\tt evclass=128}), including all four point spread function (PSF)
event types 
lying within a 15$^{\circ}$ region of interest (ROI) centered at the target source. We used a relational filter
``\texttt{DATA$\_$QUAL$>$0}'', \&\& ``\texttt{LAT$\_$CONFIG==1}'' to define good time intervals. Only the events with zenith angle of 70$^{\circ}$, 75$^{\circ}$, 85$^{\circ}$ and 90$^{\circ}$ (according to the PSF types)
were included in the analysis in order to avoid contamination from Earth-limb \gm-rays. We performed a component-wise data analysis to account for different PSF types and considered the
third catalog of \fermi-LAT detected sources \citep[3FGL;][]{2015ApJS..218...23A} to generate a source model. The source model includes all the sources present within the ROI, a Galactic 
diffuse emission component (gll\_iem\_v06.fits) and isotropic emission models (iso\_P8R2\_SOURCE\_V6\_PSF\#\_v06.txt, where \#: 0, 1, 2, and 3) \citep[][]{0067-0049-223-2-26}. A combined fitting was performed using the summed 
likelihood method included in the pyLikelihood library of the ScienceTools to derive the strength of the \gm-ray signal. This was accomplished by computing a maximum likelihood test statistic
TS =  2$\Delta \log (\mathcal{L}$) where $\mathcal{L}$ represents the likelihood function, between models with and without a point source at the position of the object \citep[][]{1996ApJ...461..396M}. 
Since we were using data below 100 MeV, we enabled the energy dispersion corrections for all sources, except for the diffuse backgrounds.
We performed a first round of optimization to obtain a best initial guess of the spectral parameters for all sources. We then allowed the spectral parameters of all the 
sources having TS$>$25 and lying within 10$^{\circ}$ from the center of the ROI to vary during the fitting. In the source spectra, only spectral bins where the source was detected with TS$>$9 are reported.

\subsection{\nustar}

The blazar 3FGL J0325.5$+$2223 was observed by \nustar~on 2015 November 8 for a net exposure of 22.2 ks; 
3FGL J0449.0$+$1121 was observed on 2015 December 2 for a net exposure of 20.5 ks; and 3FGL J0453.2$-$2808 was monitored 
on 2015 December 3 for a net exposure of 19.5 ks.
The data for both \nustar~Focal Plane Modules (FPMA and FPMB; \citealp{2013ApJ...770..103H}) were processed using the \nustar~Data Analysis Software (NUSTARDAS) v1.5.1. 
We calibrated the event data files using the task {\it nupipeline}, with the response file taken from the latest Calibration Database (CALDB).  The generation of source 
and background spectra, ancillary and response matrix files, has been achieved using the {\it nuproducts} script. 
We selected circles with radii of 30\arcsec centered on the 
target sources as the source regions and the background events were extracted from circles with the same area but from a nearby source-free region on the same frame.
\subsection{\swift}

\swift-XRT \citep[][]{2005SSRv..120..165B} and UVOT \citep[][]{2005SSRv..120...95R} observations were carried out simultaneously with \nustar~monitoring. 3FGL J0325.5$+$2223 was
observed on 2015 November 8, whereas 3FGL J0449.0$+$1121 and 3FGL J0453.2$-$2808 were monitored on 2015 December 2 and December 3, respectively. The exposure time for  each of the three targets was $\sim$2 ks. \\
Due to these short exposure times and the intrinsic faintness of the sources in this band, none of the targets were detected by UVOT. \\
The \swift-XRT observations were executed in the photon counting mode. The XRT data were analysed with the
XRTDAS software package (v.3.0.0) distributed by HEASARC within the HEASoft package 
(v.6.17). We used the task {\it xrtpipeline} to calibrate and clean the event files. Using the tool XSELECT, we extracted the source and background
regions using circles and annuli centered on the source, respectively. The radii for the two regions were chosen taking into account the difference in count 
rates for the three objects. For  3FGL J0325.5$+$2223 we used a circular region of 45\arcsec radius, and an annular region of inner radius 90\arcsec and outer radius 190\arcsec;
for  3FGL J0449.0$+$1121 we used a circular region of 12\arcsec radius, and an annular region of inner radius 40\arcsec and outer radius 140\arcsec; and for
3FGL J0453.2$-$2808 we used a circular region of 25\arcsec radius, and an annular region of inner radius 50\arcsec and outer radius 150\arcsec. \\
The ancillary response files were generated using {\it xrtmkarf} and the source spectra were rebinned to have at least one count per bin.

\subsection{GROND}

GROND is a multi-channel imager mounted on the 2.2\,m MPG\footnote{https://www.eso.org/sci/facilities/lasilla/telescopes/national/2p2.html} 
telescope at ESO in La Silla, Chile. It simultaneously observes with seven filters ({\it $g^\prime, r^\prime, i^\prime, z^\prime $}, {\it J, H, K$_s$}), covering the optical to near-infrared wavelength 
regime \citep{2008PASP..120..405G}. The data analysis procedure is described in detail in \citet{2008ApJ...685..376K}. For the optical filters ({\it $g^\prime, r^\prime, i^\prime, z^\prime $}), the point spread 
function (PSF) photometric technique was employed, whereas the aperture extraction method was applied for  the near-infrared ({\it J, H, K$_s$}) filters, because of the undersampled PSF in these bands. 
The optical filters were calibrated with the SDSS Data Release 8 \citep{2011ApJS..193...29A} and the near-IR filters were calibrated with 2MASS stars \citep{2006AJ....131.1163S}. We have corrected for Galactic extinction following 
\citet{2011ApJ...737..103S}. 
The resulting magnitudes were converted to the AB magnitude system and are provided in Table~\ref{tab:results}.

\section{Results}
\subsection{X-ray Spectral analysis}
The joint \swift-XRT (0.3-10 keV) and \nustar~(3.0-79 keV) spectra were simultaneously fitted with XSPEC using the C statistic \citep[][]{1979ApJ...228..939C}. 
For all three sources we included Galactic absorption ($N_{\rm H}$) with Galactic neutral hydrogen column densities taken from \citet[][]{2005A&A...440..775K}.

\begin{figure*}
\gridline{\fig{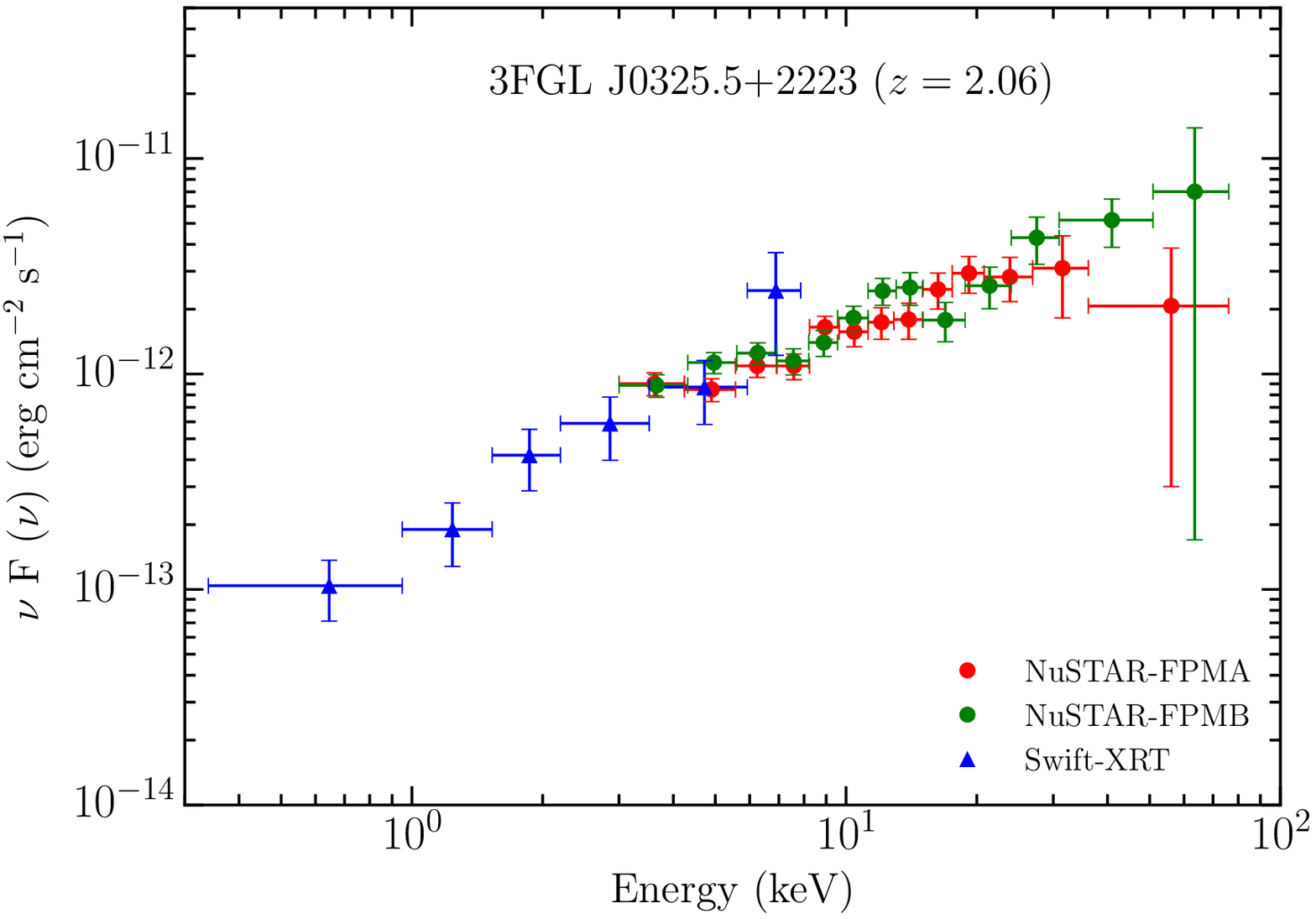}{0.5\textwidth}{(a)}
          \fig{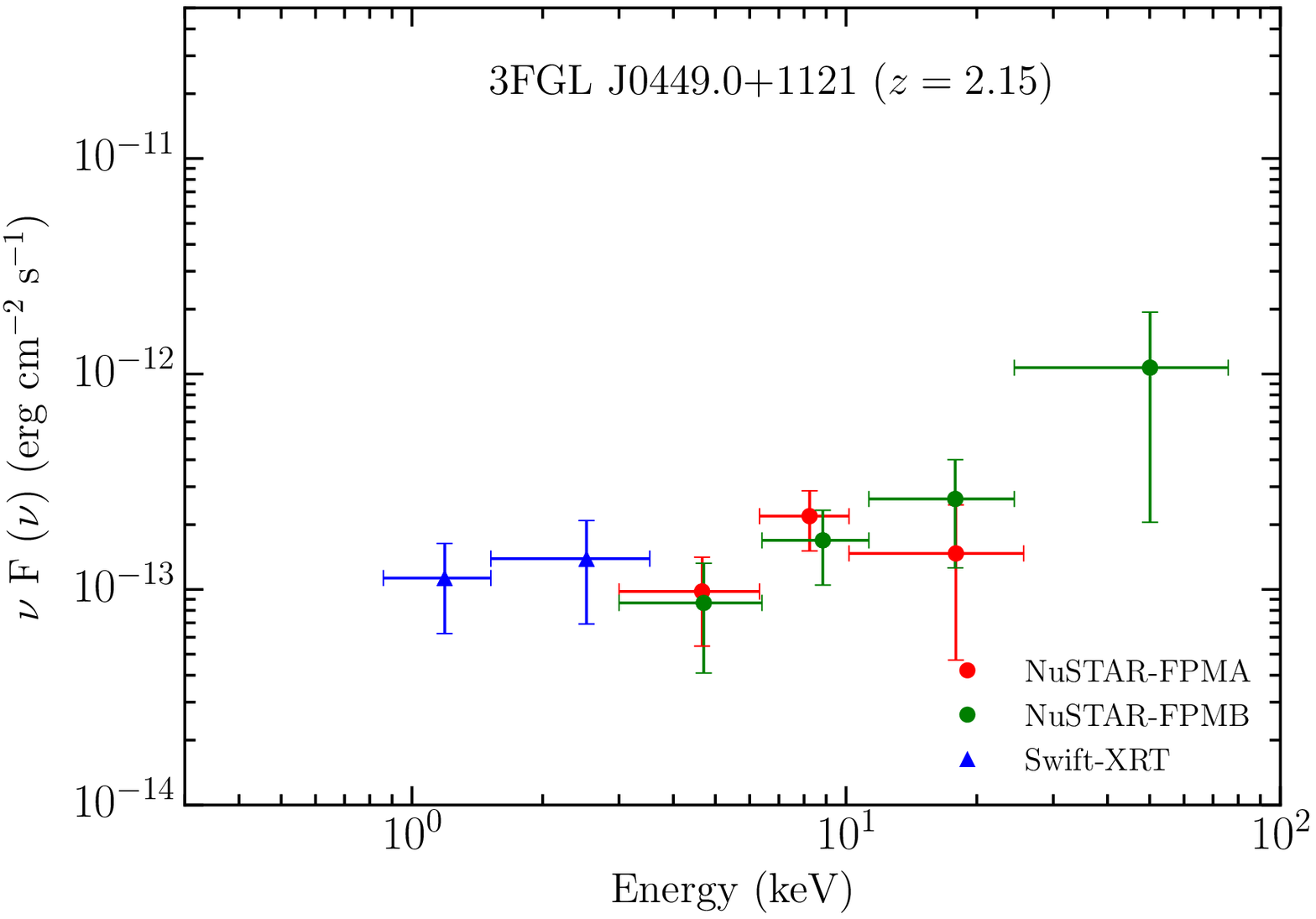}{0.5\textwidth}{(b)}
           }
\gridline{\fig{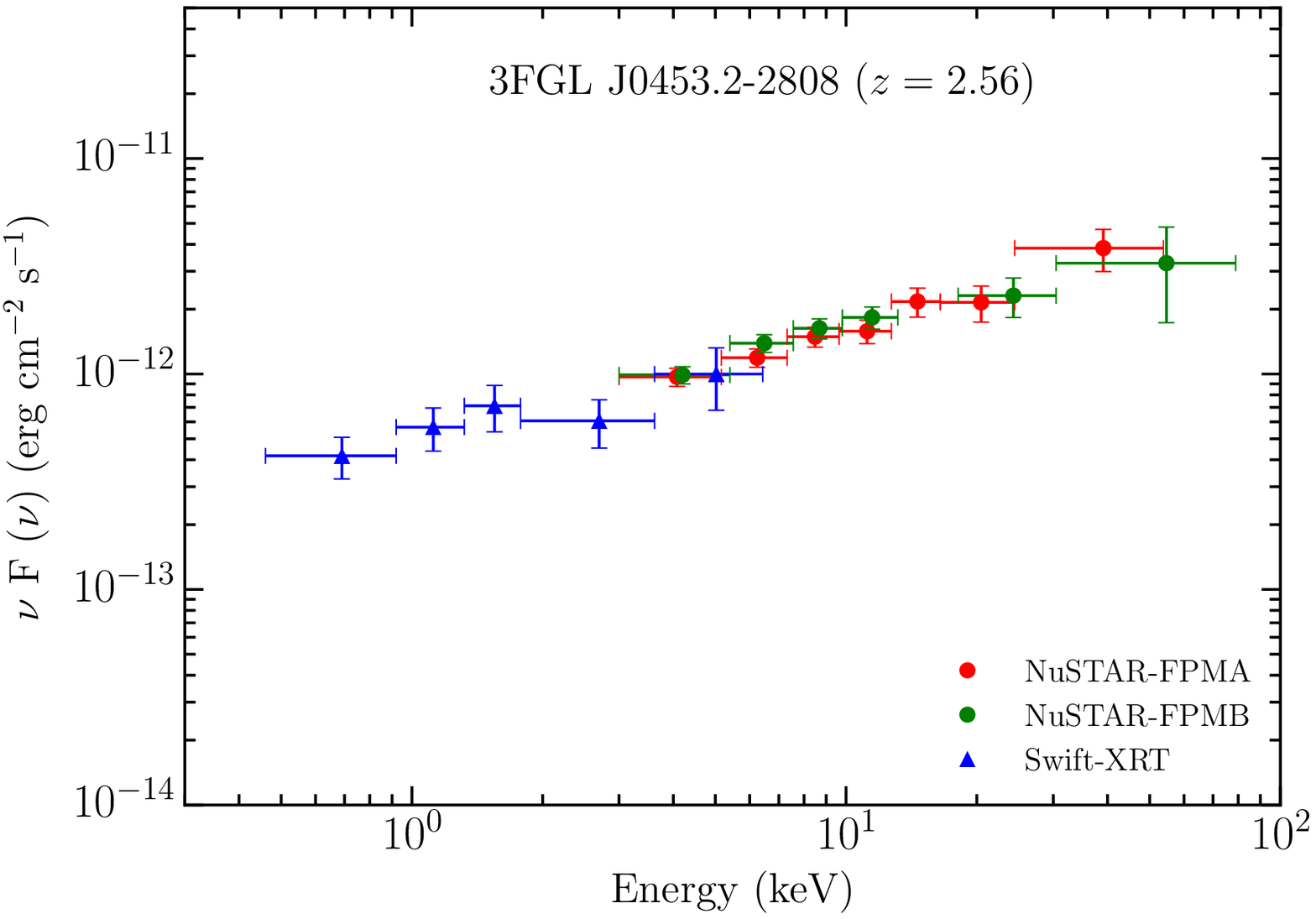}{0.5\textwidth}{(c)}
}          
\caption{Combined \swift-XRT and \nustar~(FPMA and FPMB) observations of 3FGLJ0325.5+2223 on 2015 November 11, 3FGLJ04490+1121 on 2015 
December 2, and 3FGLJ04532-2808 on 2015 December 3. 
\label{fig:combined}}
\end{figure*}

A power-law model with absorption fixed at the Galactic value was used in all three cases. We included a multiplicative constant factor to cross-calibrate the
three instruments; we kept it equal to 1 for FPMB but left it free to vary for FPMA and \swift-XRT. For two of the targets, the difference for FPMA is 
in the range of 6-7 \%, while for \swift-XRT it is in the 3-20 \% range. This is consistent with what has already been found for other sources  \citep[e.g.,][]{2015ApJS..220....8M}. In the case of 3FGL J0449.0+1121, 
due to poor photon statistics, we decided to keep the cross-calibration constant fixed to 1 for both FPMA and FPMB.
Within errors, the cross-normalization constant for \swift-XRT is compatible with 1.
The results of the spectral fits are provided in Table \ref{tab:results} and Figure \ref{fig:combined} shows, as an example, 
the combined spectrum for 3FGL J0325.5$+$2223.
 
 \subsection{SED modeling}
 To understand the underlying physical mechanisms powering the relativistic jets of these objects, we reproduced the broadband SEDs using 
 a simple one zone leptonic emission model. 
 The details of the adopted procedure can be found in \citet[][]{2009MNRAS.397..985G} and here it is briefly described.
 The observed radiation was assumed to originate from a spherical 
 emission region covering the entire cross-section of the jet, located at a distance of $R_{\rm diss}$ from the central engine, 
 and moving with the bulk Lorentz factor $\Gamma$. The jet semi opening
 angle was assumed to be 0.1 rad. The relativistic electron population was assumed to follow a broken power law energy distribution of the following type
 
 \begin{equation}
 N(\gamma)  \, \propto \, { (\gamma_{\rm break})^{-p} \over
(\gamma/\gamma_{\rm break})^{p} + (\gamma/\gamma_{\rm break})^{q}}.
\end{equation}
  where $\gamma_{\rm break}$ is the break energy, 
  {\it p} and {\it q} are the slopes of the particle energy distribution before and after $\gamma_{\rm break}$, respectively.
  In the presence of a tangled but uniform magnetic field $B$, electrons radiate via synchrotron and IC mechanisms. For the IC 
  process, the low energy photons considered are synchrotron photons and photons originating outside the jet. 
  We have considered several AGN components as potential reservoirs
  of external radiation energy density: (a) the accretion disk emission; 
  (b) the X-ray corona lying above and below the accretion disk, having a cut-off power law spectral shape, and reprocessing 30\% of the 
  disk luminosity ($L_{\rm disk}$); 
  (c) the broad line region (BLR); and (d) the dusty torus. Both the BLR and the torus are considered as spherical shells located at 
  distances $R_{\rm BLR} = 10^{17} L^{1/2}_{\rm disk,45}$ cm 
  and $R_{\rm IR} = 10^{18} L^{1/2}_{\rm disk,45}$ cm, respectively, where $L_{\rm disk,45}$ is the accretion disk luminosity in units 
  of 10$^{45}$ \lum. They are assumed to re-emit 10\% and 50\% 
  of $L_{\rm disk}$ and their spectral shapes are considered as a blackbody peaking at the Lyman-alpha frequency and $T_{\rm IR}$, respectively, 
  where $T_{\rm IR}$ is the characteristic temperature of the torus. 
  The EC spectra were calculated by deriving the comoving frame radiative energy densities from these components. Finally, we evaluated various 
  powers that the jet carries in the form of the magnetic field ($P_{\rm m}$), 
  electrons ($P_{\rm e}$), radiation ($P_{\rm r}$), and protons ($P_{\rm p}$). The $P_{\rm p}$ or kinetic jet power was estimated by
  considering protons to be cold and hence contributing only to the inertia of the jet and 
  having an equal number density to that of relativistic electrons \citep[e.g.,][]{2008MNRAS.385..283C}.
  
\begin{figure*}
\gridline{\fig{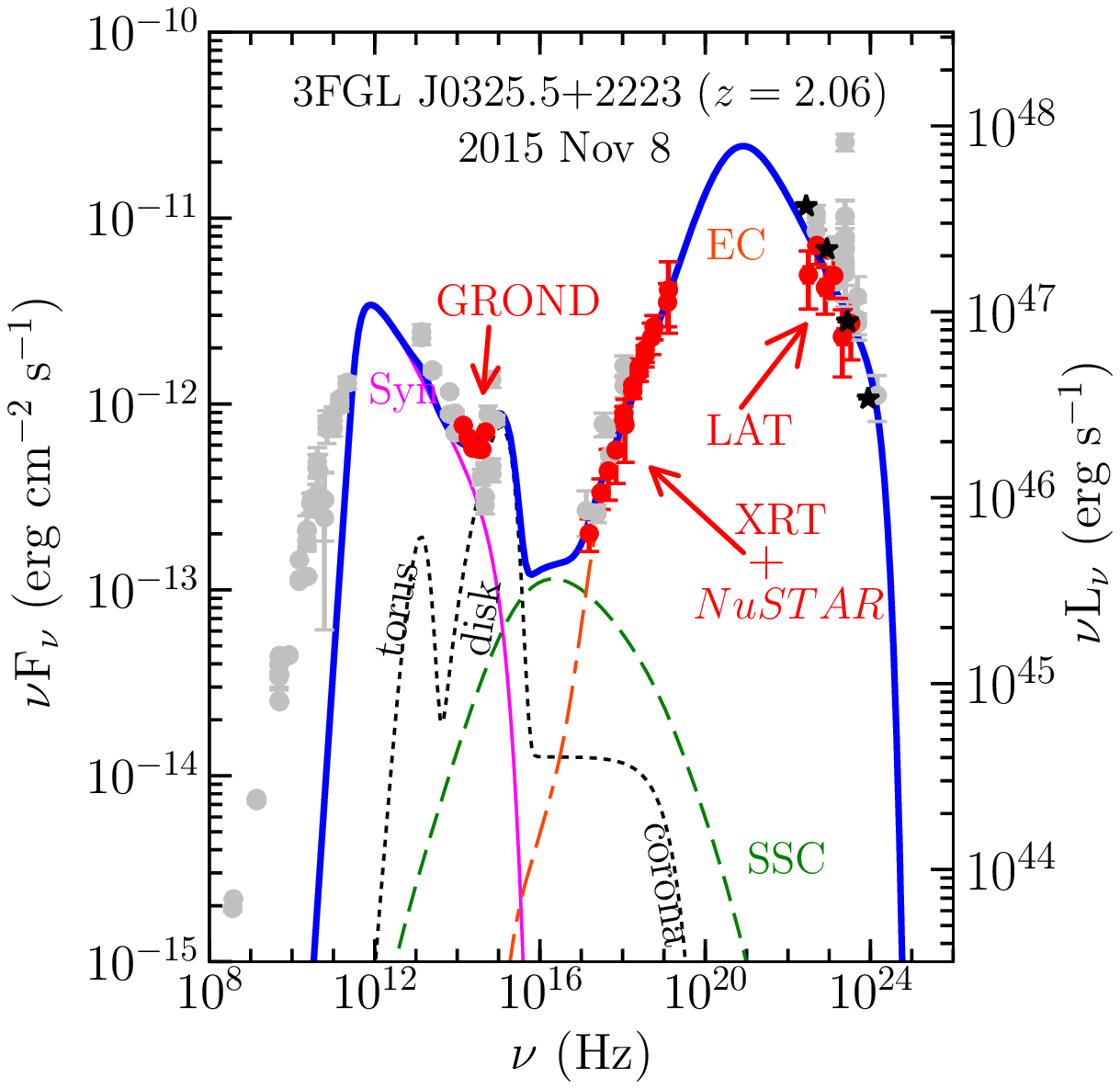}{0.5\textwidth}{(a)}
          \fig{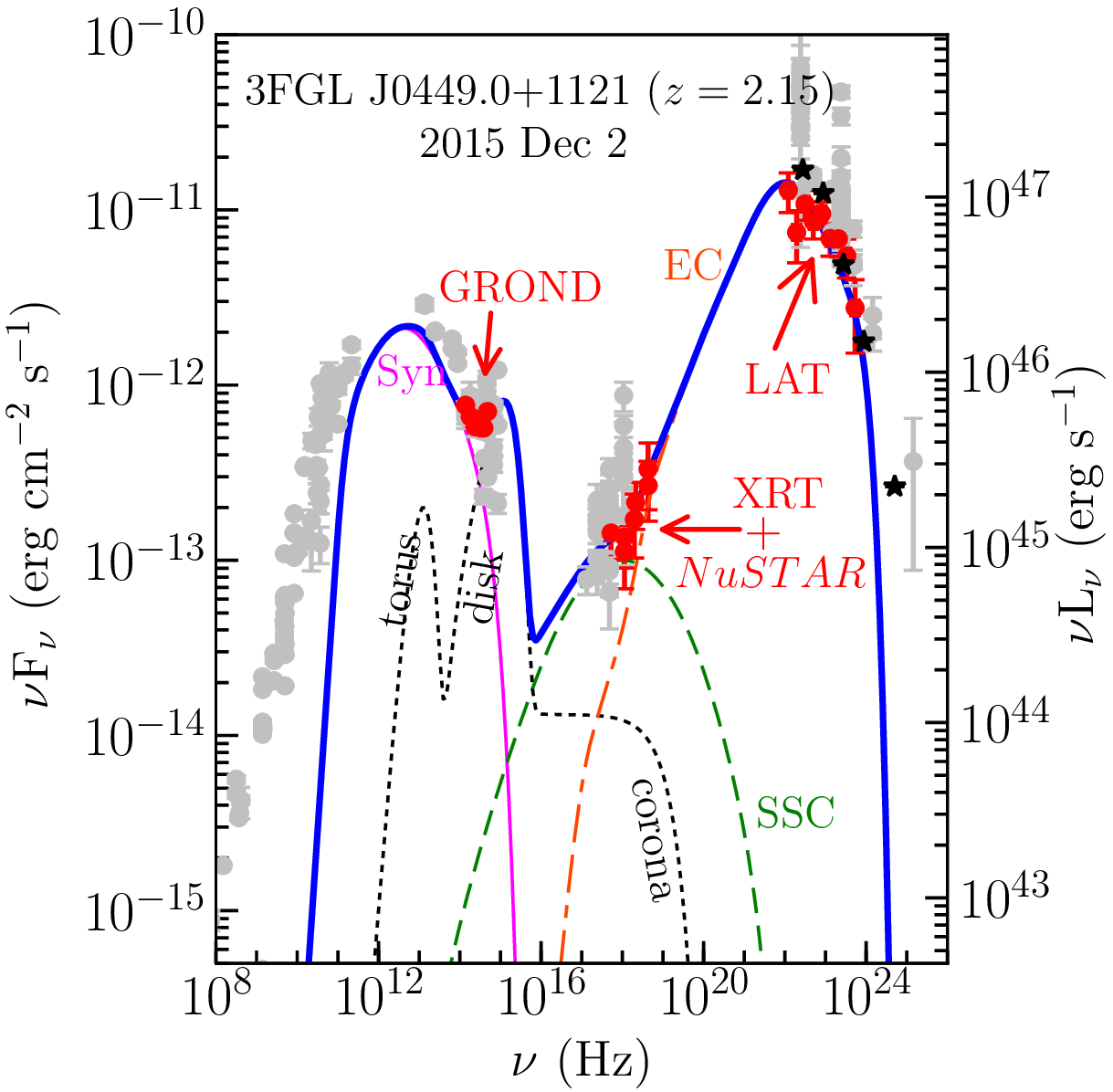}{0.5\textwidth}{(b)}
           }
\gridline{\fig{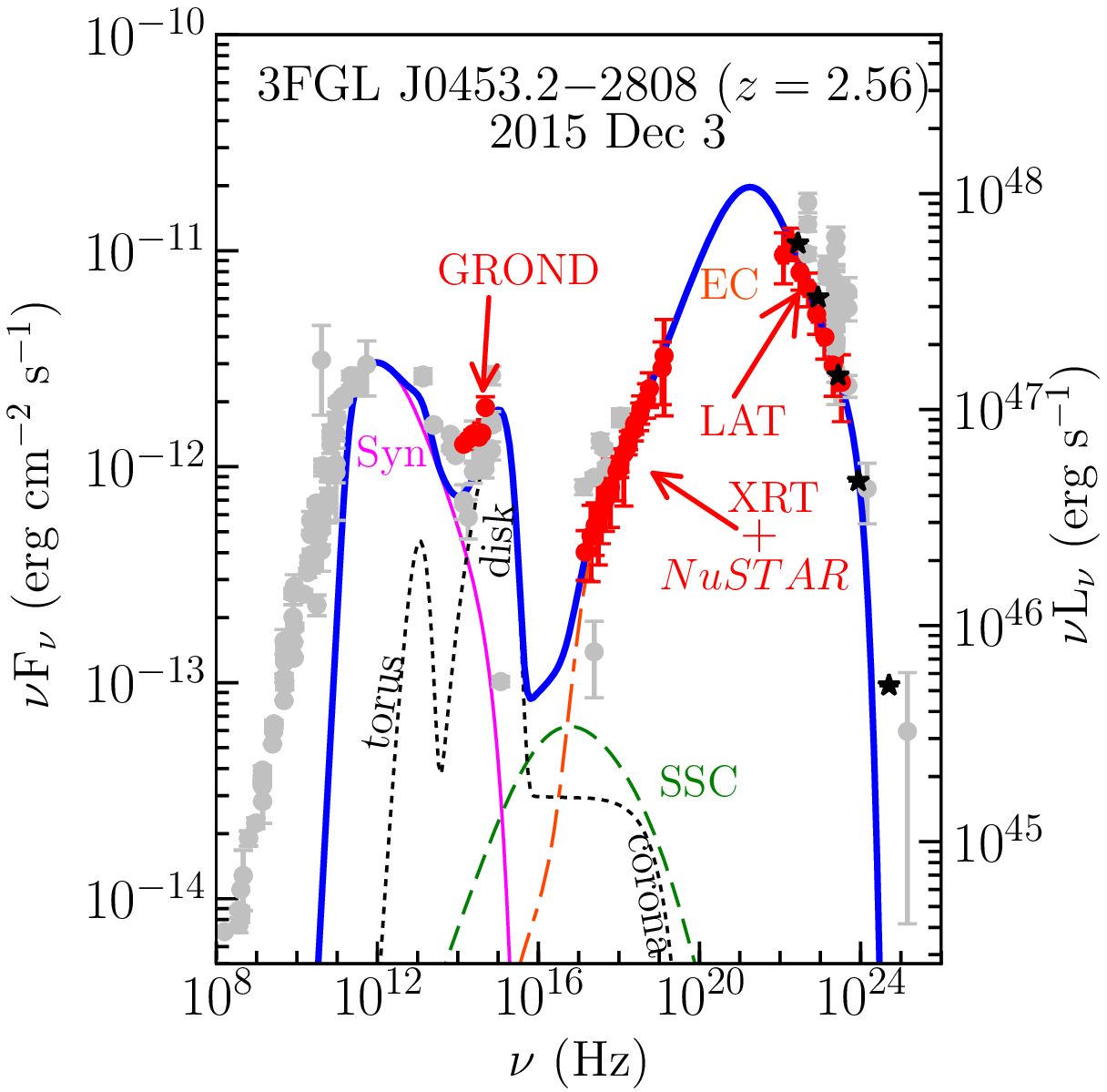}{0.5\textwidth}{(c)}
}          
\caption{The broadband SED of three quasars using quasi-simultaneous GROND, \swift, \nustar, and \fermi-LAT data, modeled using the one zone leptonic emission model described in the text. Grey and red circles represent the archival and quasi-simultaneous observations, respectively. In the \fermi-LAT energy range, black stars denote the 3FGL spectrum.\label{fig:sed}}
\end{figure*}

 The parameters associated with the SED modeling are given in Table \ref{tab:sed_par} and the results are plotted in Figure \ref{fig:sed}.

  \subsection{Black hole mass estimation}
 The most commonly used approach to calculate quasar black hole mass is adopting a single epoch optical spectroscopic measurement, which assumes that the BLR is virialized \citep[e.g.,][]{2011ApJS..194...45S}. Another novel method to derive the black hole mass is by reproducing the IR-UV spectra of quasars with a standard \citet[][]{1973A&A....24..337S}
 accretion disk, provided the big blue bump is visible in this energy band \citep[see,][]{2015MNRAS.448.1060G}. This is particularly useful when the optical/IR spectrum of the source is not available. In this technique,
 the spectral shape of the accretion disk is assumed as a multi-color blackbody with the following flux density profile \citep[][]{2002apa..book.....F}
 
 \begin{equation}
 \label{eq:SS_discflux}
 F_\nu = \nu^3\frac{4\pi h \cos \theta_{\rm v} }{c^2 {D}^2}\int_{R_{\rm in}}^{R_{\rm out}}\frac{R\,{\rm d}R}{e^{h\nu/kT(R)}-1},
\end{equation}
where $D$ is the distance of the observer, $k$ is the Boltzmann constant, $c$ is the speed of light, and $R_{\rm in}$ and $R_{\rm out}$ are the inner and outer disk radii, taken as 3$R_{\rm Sch}$ and 
500$R_{\rm Sch}$, respectively. $R_{\rm Sch}$ is the Schwarzschild radius. The radial temperature profile can be given as

\begin{equation}
T(R)\, =\, {  3 R_{\rm Sch}  L_{\rm disk }  \over 16 \pi\eta_{\rm a}\sigma_{\rm SB} R^3 }  
\left[ 1- \left( {3 R_{\rm Sch} \over  R}\right)^{1/2} \right]^{1/4},
\end{equation}
where $\sigma_{\rm SB}$ is the Stefan-Boltzmann constant and $\eta_{\rm a}$ is the accretion efficiency, adopted here as 10\%. There are only two parameters, the black hole mass and the accretion
rate $\dot{M}_{\rm a}$, to determine. The rate of accretion can be computed from the intrinsic accretion disk luminosity $L_{\rm disk}=\eta_{\rm a}\dot{M}_{\rm a}c^2$. Since $L_{\rm disk}$ can be 
obtained from observations, provided the peak of the big blue bump is visible in the SED, this leaves only the black hole mass as a free parameter \citep[e.g.,][]{2015MNRAS.448.1060G}.

The black hole masses of two out of three sources, 3FGL J0325.5$+$2223 and 3FGL J0449.0$+$1121, were derived by \citet[][]{2012ApJ...748...49S} using the optical spectroscopic approach. Using the C~{\sc iv}
line, they found masses of  $1.6 \times 10^{9}~M_{\sun}$ and $7.9 \times 10^{7}~M_{\sun}$ for 3FGL J0325.5$+$2223 and 3FGL J0449.0$+$1121, respectively. For 3FGL J0453.2$-$2808, we used C~{\sc iv} 
line parameters from \citet[][]{1983A&A...117...60F} and adopted the empirical relations of \citet[][]{2011ApJS..194...45S} to derive a central black hole mass of $\sim7.9 \times 10^{8}~M_{\sun}$. Instead following the SED 
modeling approach, we found black hole masses for 3FGL J0325.5$+$2223, 3FGL J0449.0$+$1121 and 3FGL J0453.2$-$2808 of $6.3 \times 10^{8}$, $5.0 \times 10^{8}$ and $1.0 \times 10^{9}~~M_{\sun}$, respectively (all black-hole
masses are listed in Table \ref{tab:mass}). The black hole masses computed from both approaches reasonably match within a factor of 2 for two of the sources\footnote{It should be noted that the typical errors in virial spectroscopic 
black hole mass calculation is $\sim$0.4 dex \citep[e.g.,][]{2006ApJ...641..689V,2011ApJS..194...45S}}, though the SED modeling predicts a higher black hole mass (factor of $\sim$6) for 3FGL J0449.0+1121.

 \section{Discussion}
 High-redshift blazars are bright targets in hard X-rays. This is probably due to the shift of blazar SED towards longer
wavelengths as their non-thermal luminosity increases \citep{1999ASPC..159..351F}.
 The shifting causes their spectra to become steeper at $\gamma$-rays 
 ($\Gamma_{\gamma}\gtrsim 2.3$) and harder at X-rays 
($\Gamma_{\rm X}\lesssim 1.5$). Indeed, all the three MeV blazars studied here
display these features (see Table \ref{tab:results}). 
In this regard, observations in both these energy bands are crucial to determine the power of the jet and explore its
connection to accretion. In fact, having both \nustar~ and \fermi-LAT detections for all three sources provides a unique opportunity to locate the IC peak and study the shape of the underlying electron
population. The bolometric emission in such powerful blazars is dominated by high energy X-ray and \gm-rays radiation conveying that good quality spectral measurements in both bands are desirable.
Furthermore, there are a few other interesting properties of high redshift blazars 
revealed by \nustar~ monitoring. This includes a spectral 
flattening seen in the joint \swift-XRT and \nustar~ spectrum of various MeV blazars \citep{2016ApJ...825...74P}
and also a substantial flux variability seen at two different epochs of \nustar~
monitoring \citep[e.g.,][]{2015ApJ...807..167T}. The latter becomes more important due to the fact that
these sources are weaker at $\gamma$-rays and may not have enough signal to detect significant variability
in this energy range. As a result, \nustar~ has proved to be a fundamental instrument
to pursue high-redshift blazar studies.

\begin{figure*}

\centering
\includegraphics[width=0.6\textwidth]{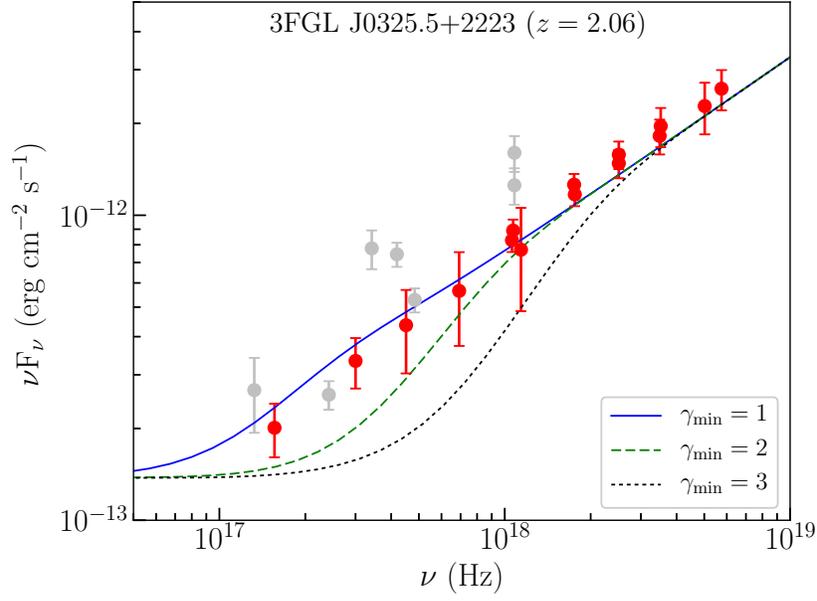}
\caption{ Zoomed SED of 3FGL J0325.5$+$2223, showing the X-ray spectrum. The different lines represent the modeling done with various
$\gamma_{\rm min}$ values (as labelled). As can be seen, in this source the low-energy cut-off
cannot be significantly larger than unity.\label{fig:gamma_min}}
\end{figure*}

The joint XRT and NuSTAR spectra of the three sources 
do not show any curvature within the available statistics and is well fitted by a simple absorbed power law model
with $N_{\rm H}$ fixed to the Galactic value. It is reported in various
recent studies that the 0.3-79 keV X-ray spectra of many MeV blazars ($>5$) show 
a distinct curvature or a break around $\sim$few keV \citep[e.g.][]{2015ApJ...807..167T,2016ApJ...825...74P}. 
Such a feature reflects the behavior of the emitting electron distribution, 
intrinsic to the jet, rather than any other external factors \citep[][and 
references therein]{2016ApJ...825...74P}. On the other hand, there are observations of MeV blazars, 
that do not show any such feature \citep[e.g.][]{2016ApJ...826...76A}. In fact, the shape of the X-ray spectrum
constrains the behavior of the underlying electron population, especially the low energy cut-off of the electrons
($\gamma_{\rm min}$) provided the X-ray emission is dominated by EC process, rather than by SSC \citep[see, e.g.,][]{2008MNRAS.385..283C}.
This is illustrated in Figure \ref{fig:gamma_min}. As can be seen, a good fit can be achieved only with $\gamma_{\rm min}\sim 1$. For higher
values, the model predicts a significant break in the X-ray spectrum, which is not seen\footnote{We caution that our calculations at low values
of $\gamma$ are rather approximate. These relatively `cold' electrons could be involved in bulk Compton process \citep[][]{2007MNRAS.375..417C} and 
one should self-consistently take this into account. However, this feature is yet to be observed \citep[however see,][for a possible detection]{2012ApJ...751..159A} and so its contribution is uncertain. Furthermore, 
the kinetic jet power is very sensitive to $\gamma_{\rm min}$ and for $\gamma_{\rm min}\gg1$ it falls below the jet power in radiation, 
which is problematic for powerful FSRQs \citep[see, e.g.][]{2014Natur.515..376G}}. 
This implies the joint XRT and {\it NuSTAR} observations
are instrumental in evaluating the minimum energy of the underlying particle population, which ultimately controls the jet power.

 The broadband SEDs of the three MeV sources resemble those of powerful blazars. The synchrotron peaks lie in the sub-mm range, 
 whereas the high-energy IC peaks lie in the MeV band. The Compton dominance 
 (the ratio of IC to synchrotron peak luminosities) is found to be $>$1 for all three blazars.
 The GROND observations reveal a break
 in the IR-optical spectra of the sources which we interpret as a combination of the falling synchrotron 
 spectrum and the rising accretion disk radiation. Though we do not see the peak of the disk emission 
 (primarily due to lack of UV data), based on the shape of the GROND spectra and the available archival
 broad line luminosities \citep[][]{1983A&A...117...60F,2012ApJ...748...49S}, we are able to derive both the disk luminosity and the 
 central black hole mass. Another constraint is provided by the broader limiting range of the disk luminosity that can be set by 
 considering 10$^{-2}L_{\rm Edd}<L_{\rm disk}<L_{\rm Edd}$. The upper limit ensures the source to be sub-Eddington and the lower limit assumes the accretion disk is radiatively efficient. Furthermore,
 combining the observations from \swift-XRT, \nustar~ and \fermi, we can cover the X- and \gm-ray portions of the spectrum.
 These parts of the SEDs are successfully reproduced by IC scattering of low energy photons 
 produced externally to the jet. 
 
 Since in the adopted model the radiative energy densities of various AGN components are 
 functions of the distance from the central engine, we are able to determine the location 
 of the emission region. This is found to be at the outer edge of the BLR where a major fraction of the seed photons are supplied by
 the BLR clouds. The X-ray spectra of all three objects are extremely hard and 
 observations further hint at the dominance of the EC process over SSC emission in this energy range \citep[see,][for a detailed 
 discussion]{2016ApJ...826...76A}. On the other hand, the \gm-ray SEDs are steep, 
 as expected in high redshift blazars. The spectral shapes observed at X-ray and \gm-ray energies help constrain the slopes of the 
 underlying broken power law electron energy distribution. Overall, 
 the observed SED parameters of these three sources are similar to other high-redshift MeV blazars \citep[e.g.,][]{2010MNRAS.405..387G}.
 
\begin{figure*}

\centering
\includegraphics[width=0.9\textwidth]{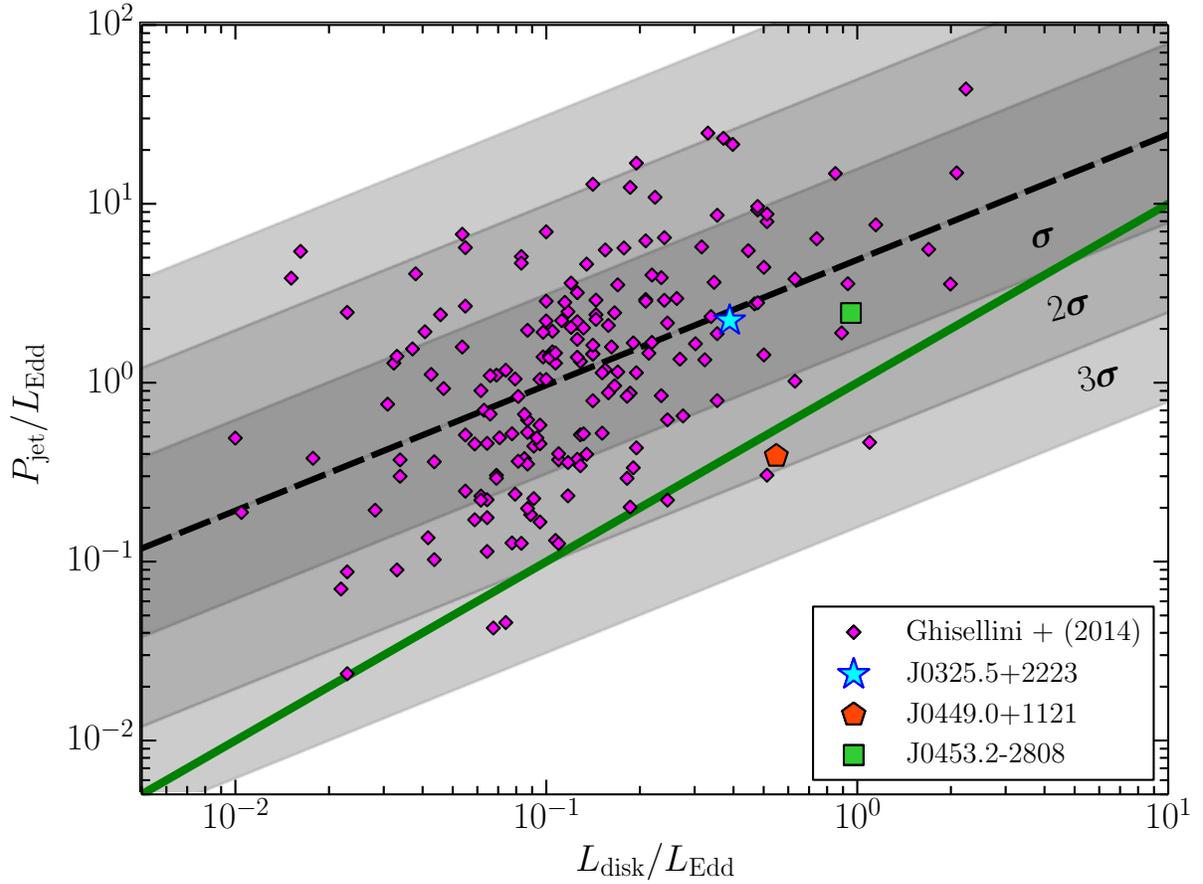}
\caption{Comparison of jet powers and accretion disk luminosities of three MeV sources with that of a large sample of blazars studied by \citet[][]{2014Natur.515..376G}.
Black dashed and green solid lines represent the best fit and one-to-one correlation of the plotted quantities, respectively. Both jet powers and
disk luminosities are normalized for central black hole mass.\label{fig:comparison}}
\end{figure*}

 With good quality IR-optical (constraining the accretion disk emission) and 
 hard X-ray-\gm-ray data (required for the accurate measurement of the jet power) in hand,
 it is interesting to compare the disk-jet connection observed in the three
 MeV blazars studied in this work with that for other blazars.
 With this in mind, we collect the jet powers and disk luminosities of
 all blazars studied by \citet[][]{2014Natur.515..376G}. In Figure \ref{fig:comparison}, we plot the jet power as a function of
 the disk luminosity for blazars, including our three MeV sources.
 The plotted quantities are normalized for the central black hole mass. As can be seen, the majority of sources have normalized jet power
 exceeding their
 normalized disk luminosities (the one-to-one correlation is shown by the green solid line). 3FGL J0325.5+2223
 and 3FGL J0453.2$-$2808 have their jet powers larger than their disk 
 luminosities, though 3FGL J0449.0+1121 lies just below the one-to-one correlation. Indeed, among the three objects, this object has the
 least jet power. The jet powers of both  3FGL J0325.5+2223 and 
 3FGL J0453.2$-$2808 appear to be larger than the Eddington luminosity ($P_{\rm jet}/L_{\rm Edd}>1$).
 However, there are a 
 few caveats. 
 The existence of electron-positron pairs in the jet would reduce the jet power 
 by $n_{\rm e}$/$n_{\rm p}$ \citep[where $n_{\rm e}\equiv n_{e^-} + n_{e^+}$, see,][for a detailed discussion]{doi:10.1093/mnras/stw2960}, 
 although their number could not greatly exceed the protons in order to avoid the Compton rocket effect that would 
 otherwise stop the jet \citep[][]{2010MNRAS.409L..79G}. Furthermore, the budget of the jet power can also come down if one considers a spine-sheath structured jet,
 instead of a one-zone uniform emission region \citep[][]{2016MNRAS.457.1352S}. Most importantly, the underestimation 
 of the black hole mass can also lead to super-Eddington jet power. Consequently,
 one has to take into account these uncertainties and/or their combination
 before drawing any firm conclusion.

\section{Conclusions}
We studied three high redshift ($z>2$) MeV blazars using quasi-simultaneous GROND, \swift, \nustar, and \fermi-LAT data, focusing 
on the hard X-ray part uncovered by \nustar. With the latter and the aid of the \fermi-LAT data we were able to constrain the position of the IC peak and to derive related characteristics 
of these sources like bolometric luminosity, jet power as well as the relativistic particle distribution and the location of the emission region. Our primary findings are as follows
\begin{enumerate}
\item All sources are significantly detected by \nustar~and exhibit a flat (photon index $\lesssim1.5$) X-ray spectrum extended up to 79 keV, as revealed from joint XRT and \nustar~fitting.
\item The broadband SEDs of these sources resembles that of powerful blazars with synchrotron and IC peaks lying at sub-mm and MeV energy ranges,
respectively.
\item  The IR-optical spectra can be explained by a combination of synchrotron and accretion disk spectrum, whereas, 
high energy X-ray to \gm-ray radiation is successfully reproduced 
by IC scattering of low energy photons mainly supplied by the BLR.
\item The location of the emission region is found to be at the outer edge of the BLR in all three sources. The black hole masses for all three sources are greater than $10^{8.6}~M_{\sun}$. 
\item Comparing the normalized jet powers and disk luminosities of these sources with 
 that of a large sample of blazars, we find them to lie on the well known 
 disk-jet correlation derived by \citet[][]{2014Natur.515..376G} where their jet powers
 exceeds accretion disk luminosities. Only 3FGL J0449.0+1121 shows a slight offset 
 from this correlation. 
\end{enumerate}
 
\acknowledgements
We are thankful to the referee for a constructive report.
The \textit{Fermi} LAT Collaboration acknowledges generous ongoing support
from a number of agencies and institutes that have supported both the
development and the operation of the LAT as well as scientific data analysis.
These include the National Aeronautics and Space Administration and the
Department of Energy in the United States, the Commissariat \`a l'Energie Atomique
and the Centre National de la Recherche Scientifique / Institut National de Physique
Nucl\'eaire et de Physique des Particules in France, the Agenzia Spaziale Italiana
and the Istituto Nazionale di Fisica Nucleare in Italy, the Ministry of Education,
Culture, Sports, Science and Technology (MEXT), High Energy Accelerator Research
Organization (KEK) and Japan Aerospace Exploration Agency (JAXA) in Japan, and
the K.~A.~Wallenberg Foundation, the Swedish Research Council and the
Swedish National Space Board in Sweden. Additional support for science analysis during the operations phase is gratefully 
acknowledged from the Istituto Nazionale di Astrofisica in Italy and the Centre National d'\'Etudes Spatiales in France.

This \nustar~work was supported under NASA Contract No. NNG08FD60C, and made
use of data from the \nustar~mission, a project led by the California Institute of Technology,
managed by the Jet Propulsion Laboratory, and funded by the National Aeronautics and
Space Administration. We thank the \nustar~Operations, Software and Calibration teams
for support with the execution and analysis of these observations. This research has made
use of the \nustar~Data Analysis Software (NuSTARDAS) jointly developed by the ASI
Science Data Center (ASDC, Italy) and the California Institute of Technology (USA).

We thank the \swift~team and the Swift PI (N. Gehrels) for promptly scheduling and executing the observations.

Part of this work is based on archival data, software or on–line services provided by the
ASI Data Center (ASDC). This research has made use of the XRT Data Analysis Software
(XRTDAS). Part of the funding for GROND (both hardware and personnel) was generously
granted by the Leibniz-Prize to G. Hasinger (DFG grant HA 1850/28-1).

We acknowledge funding from NASA contract NNX15AV09G. M.\,B. acknowledges support from NASA Headquarters under the NASA Earth and Space Science Fellowship Program, grant NNX14AQ07H.

Part of the funding for GROND (both hardware as well as personnel) was generously granted from the Leibniz-Prize to Prof. G. Hasinger (DFG grant HA 1850/28-1).

\newpage

\begin{table*}[t!]
\scriptsize
 \begin{center}
 \caption{Table of Observations and Spectral Parameters}\label{tab:results}
 \resizebox{\textwidth}{!}{
 \begin{tabular}{ c c c c c c c c c } 
 \hline
 \hline
  \multicolumn{9}{c}{\fermi-LAT} \\
 \hline
  Name & Time Covered & Flux\tablenotemark{a} & Photon Index\tablenotemark{b} & Test Statistic\tablenotemark{c} & & & & \\
  \hline
  3FGL J0325.5$+$2223 & 2015-03-01$-$2016-03-01 & $5.57\pm{1.12}$ & $2.37\pm{0.09}$ & 76.36 & & & &\\
  3FGL J0449.0$+$1121 & 2015-03-01$-$2016-03-01 & $11.90\pm{1.50}$ & $2.35\pm{0.06}$ & 229.62 & & & &\\
  3FGL J0453.2$-$2808 & 2015-03-01$-$2016-03-01 & $8.11\pm{1.22}$ & $2.49\pm{0.08}$ & 195.91 & & & &\\
 \hline
 \multicolumn{9}{c}{\nustar~$+$ \swift-XRT} \\
 \hline
  Name & Date & $N_{\rm H}$ \tablenotemark{d} & Photon Index \tablenotemark{e}& Flux \tablenotemark{f} & C-Stat/dof & & & \\  
  \hline
  3FGL J0325.5$+$2223 & 2015-11-08 & $8.92$ & $1.36^{+0.10}_{-0.09}$ & $7.44^{+0.81}_{-1.13}$ & 643.74/823 & & & \\
  3FGL J0449.0$+$1121 & 2015-12-02 & $12.6$ & $1.46^{+0.44}_{-0.43}$ & $1.05^{+0.40}_{-0.91}$ & 291.47/331 & & & \\
  3FGL J0453.2$-$2808 & 2015-12-03 & $2.05$ & $1.52 \pm{0.10}$ & $6.64^{+0.91}_{-0.70}$ & 625.66/769 & & & \\
  \hline
  \multicolumn{9}{c}{GROND} \\
  \hline
  Name & {UT Date\tablenotemark{g}} & \multicolumn{7}{c}{AB Magnitude\tablenotemark{h}} \\
  &  & $g^\prime$ & $r^\prime$ & $i^\prime$ & $z^\prime$ & $J$ & $H$  & $K_s$ \\
  \hline
  3FGL J0325.5$+$2223 &  2015-11-15.49 & $ 18.91\pm 0.05$ & $ 18.78\pm 0.04$ &  $18.58\pm 0.05$  &  $18.25 \pm 0.05$ & $ 18.36\pm 0.11$ & $ 18.40\pm 0.13$ & $ 17.91\pm 0.16$  \\
  3FGL J0449.0$+$1121     & 2015-12-08.31& $ 21.17\pm 0.05$ & $ 18.49\pm 0.05$ & $ 18.49\pm 0.04$ & $ 18.32\pm 0.05$ & $ 17.94\pm 0.12$ & $ 17.50\pm 0.14$ & $ 17.05\pm 0.14$\\
  3FGL J0453.2$-$2808   & 2016-06-01.63 & $ 17.35\pm 0.29$ & $ 17.42\pm 0.30$ & $ 17.49\pm 0.29$ & $ 17.37\pm 0.28$ & $ 16.96\pm 0.10$ & $ 16.72\pm 0.12$ & $ 16.49\pm 0.13$ \\
  \hline
\multicolumn{9}{l}{%
  \begin{minipage}{\textwidth}
 \tablenotetext{} {\bf Notes:}
 \tablenotetext{a}{Integrated \gm-ray flux in the $0.06-300$ GeV energy range in units of $10^{-8}$ photons {\mbox{${\rm \, cm^{-2} \, s^{-1}}$}}.}
 \tablenotetext{b}{Photon index calculated from \gm-ray analysis.}
 \tablenotetext{c}{Test statistic is a measure of significance of detection \citep[$\sigma \sim \sqrt{\rm TS}$;][]{1996ApJ...461..396M}.}
 \tablenotetext{d}{Column density in units of $10^{20}$ {\mbox{${\rm \, cm^{-2}}$}}.}
 \tablenotetext{e}{Photon index calculated from X-ray analysis.}
 \tablenotetext{f}{Observed flux in units of $10^{-12}$\ergflux~in the $0.3-79$ keV energy band. The errors are at the 90\% level of confidence for one parameter of interest and the fluxes are corrected for the Galactic absorption.}
 \tablenotetext{g}{Exposure start time.} 
 \tablenotetext{h}{Corrected for Galactic reddening.}
\end{minipage}%
}\\
 \end{tabular}}
\end{center}

\end{table*}

\begin{table*}[t!]
\begin{center}
\caption{Summary of the parameters used/derived from the SED modeling of three MeV blazars shown in Figure~\ref{fig:sed}. A viewing angle of 3$^{\circ}$ is adopted for all of them.}\label{tab:sed_par}
\begin{tabular}{lccccc}
\tableline
Parameter 									  & J0325.5+2223 & J0449.0+1121 & J0453.2$-$2808 \\
\tableline
\tableline
Slope of the particle distribution below the break energy ($p$)                     & 1.45                 & 1.55             & 1.95    \\
Slope of the particle distribution above the break energy ($q$)                       & 3.9                   & 4.15             & 4.1     \\
Magnetic field in Gauss ($B$)                                                                       & 3.2                  & 1.5               & 2.5     \\
Particle energy density in erg cm$^{-3}$ ($U_{e}$)                                     & 0.03              & 0.01                & 0.01 \\ 
Bulk Lorentz factor ($\Gamma$)                                                                  & 10                   & 12                & 10      \\
Minimum Lorentz factor ($\gamma_{\rm min}$)                                          & 1                     & 6                  & 1        \\
Break Lorentz factor ($\gamma_{\rm break}$)                                                  & 57                   & 310              & 139      \\
Maximum Lorentz factor ($\gamma_{\rm max}$)                                         & 3e3                 & 3e3              & 3e3    \\
Dissipation distance in parsec ($R_{\rm Sch}$)                                             & 0.18 (3090)     & 0.23 (4850) & 0.37 (3900) \\
Size of the BLR in parsec (in $R_{\rm Sch}$)                                                & 0.18 (3091)     & 0.19 (4006) & 0.35 (3709) \\
Black hole mass in log scale, in units of solar mass  ($M_{\rm BH,m}$)           & 8.8                 & 8.7           & 9.0   \\
Accretion disk luminosity in log scale ($L_{\rm disk}$, erg s$^{-1}$)             & 46.48               & 46.54         & 47.08 \\
Accretion disk luminosity in Eddington units ($L_{\rm disk}/L_{\rm Edd}$)    & 0.40                 & 0.56           & 0.95   \\
Characteristic temperature of IR-torus in Kelvin ($T_{\rm IR}$)             & 500     & 500    &  500 \\
Observed variability time scale in days ($t_{\rm var}$)                        & 4          &  5      &  10 \\
\hline
Jet power in electrons in log scale ($P_{\rm e}$, erg s$^{-1}$)                     & 44.87               & 44.52        & 44.90 \\
Jet power in magnetic field in log scale ($P_{\rm B}$), erg s$^{-1}$              & 46.06               & 45.79        & 46.49 \\
Radiative jet power in log scale ($P_{\rm r}$, erg s$^{-1}$)                          & 45.74               & 45.60        & 45.92 \\
Jet power in protons in log scale ($P_{\rm p}$, erg s$^{-1}$)                        & 47.18              & 46.15         & 47.43 \\
\tableline
\end{tabular}
\end{center}
\end{table*}

\begin{table*}[t!]
\scriptsize
 \begin{center}
 \caption{Table of Black-Hole Masses, derived both from spectroscopic approach and SED modelling.}\label{tab:mass}
 \begin{tabular}{ c c c c } 
 \hline
 & 3FGL J0325.5$+$2223 & 3FGL J0449.0$+$1121 &  3FGL J0453.2$-$2808 \\
 \hline
 \hline
 $M_{\rm BH, SED}$ ($M_{\sun}$) &  $6.3 \times 10^{8}$ & $5.0 \times 10^{8}$ & $1.0 \times 10^{9}$ \\
 \hline
 $M_{\rm BH, spectroscopy}$ ($M_{\sun}$) & $1.6 \times 10^{9}$ & $7.9 \times 10^{7}$ & $7.9 \times 10^{8}$ \\
 \hline
\end{tabular}
\end{center}
\end{table*}

\bibliographystyle{aasjournal}
\bibliography{MeV}

\end{document}